\documentclass[twoside,twocolumn,fleqn,superscriptaddress,showkeys]{revtex4-2}
\usepackage{xcolor}    
\usepackage{graphicx}  
\usepackage{amssymb}   
\usepackage{amsmath}   
\usepackage{bm}        
\usepackage{times}     
\usepackage{fancyhdr}  
\usepackage{float}
\newcommand{\la}{\left\langle}
\newcommand{\ra}{\right\rangle}
\newcommand{\be}{\begin{equation}}
\newcommand{\ee}{\end{equation}}
\newcommand{\bse}{\begin{subequations}}
	\newcommand{\ese}{\end{subequations}}
\newcommand{\bea}{\begin{eqnarray}}
\newcommand{\eea}{\end{eqnarray}}
\newcommand{\ba}{\begin{array}}
	\newcommand{\ea}{\end{array}}

\begin{document} 

\title{Equilibrium and nonequilibrium properties of Euler turbulence} 
\author{Mahendra K Verma} 
\affiliation{Department of Physics, Indian Institute of Technology Kanpur, Kanpur 208016, India} 
\email{mkv@iitk.ac.in}

\author{Soumyadeep Chatterjee}
\affiliation{Department of Physics, Indian Institute of Technology Kanpur, Kanpur 208016, India} 
\email{inspire.soumya@gmail.com}

\author{Shashwat Bhattacharya}
\affiliation{Institute for Thermodynamics and Fluid Dynamics, Technische Universität Ilmenau, Postfach 100565, D-98684 Ilmenau, Germany} 


\date{\today} 

\begin{abstract}
In this article, we report the equilibrium and nonequilibrium features of two-dimensional (2D) and three-dimensional (3D) Euler turbulence. To obtain a full range of equilibrium spectra, we perform pseudo-spectral simulations of Euler turbulence using $\delta$-correlated velocity field as an initial condition. These simulations provide zero energy flux and Maxwell-Boltzmann distribution for the velocity field, thus providing  direct verification of the absolute equilibrium theory of turbulence.  However, for ordered initial condition, 2D Euler turbulence remains out of equilibrium, with flow getting more ordered with time. We show that the  hydrodynamic entropy of 2D Euler turbulence decreases with time, even though the system is isolated. 
\end{abstract}

\keywords{Arrow of time, Nonequilibrium phenomena, Energy transfers, Multiscale physics, Turbulence.}

\maketitle  
\thispagestyle{fancy}  


\section{INTRODUCTION} 

The present article is based on an invited talk delivered in APPC15. It contains review of past works on Euler turbulence, as well as several new results, e.g., nonequilibrium behaviour of two-dimensional (2D) Euler turbulence, and hydrodynamic entropy.

Physical processes are either in equilibrium  or out of equilibrium~\cite{Zwanzig:book,Livi:book}.  Thermodynamics provides many examples of equilibrium processes, e.g., thermal gas, Bose gas, magnetic systems in heat bath.  In a gas or liquid under equilibrium, there is no net flow of energy or matter from one region to another statistically.  This property is called {\em detailed balance}.    Besides, the average energy and entropy of an equilibrium system remain invariant in time~\cite{Landau:book:StatMech,Reif:book:StatMech}.

On the other hand, nonequilibrium systems are time-dependent with detailed balance broken~\cite{Zwanzig:book,Livi:book}.   Earth's atmosphere, turbulent convection, hydrodynamic turbulence, and earthquakes are examples of such systems.    Kolmogorov's theory of turbulence~\cite{Kolmogorov:DANS1941Dissipation,Kolmogorov:DANS1941Structure,Frisch:book,Lesieur:book:Turbulence} provides valuable insights into  dissipative hydrodyanmic turbulence.  In this theory,  a viscous fluid is forced at large scales.   The energy injected at the large scale  is transferred to intermediate scale (called inertial range) and then to small scales, where the injected energy is dissipated.  Under a steady state, the inertial-range  energy spectrum is $E(k) = K_\mathrm{Ko} \epsilon^{2/3} k^{-5/3}$, where $ K_\mathrm{Ko}$ is Kolmogorov's constant, $\epsilon$ is the energy flux in the inertial range, and $k$ is the wavenumber.  

In this paper, we focus on turbulence  in incompressible Euler equation, which is the hydrodynamic equation with zero  external force and zero viscosity.  As we describe below, turbulence in Euler equation, referred to as Euler Turbulence, is very different from  Kolmogorov's model of turbulence, which applies to viscous flows.  Kraichnan~\cite{Kraichnan:JFM1973} and Lee~\cite{Lee:QAM1952} argued that  Euler turbulence has similarities with \textit{equilibrium thermodynamics}, and constructed \textit{absolute equilibrium theory} of Euler turbulence.  By  invoking Liouville's theorem for the Fourier modes of Euler equation, Kraichnan and Lee derived equilibrium solution of the   three-dimensional (3D) Euler equation with a  finite number of Fourier modes, also called \textit{truncated Euler equations}. For this solution, the kinetic energy flux vanishes, and the kinetic energy spectra are
 \bea
E(k) \propto \frac{k^2}{\beta - \gamma k^2}~\mathrm{for~3D},
\label{eq:Euler3D} \\
E(k)  \propto \frac{k}{\beta + \gamma k^2}~\mathrm{for~2D},\label{eq:Euler2D}
\eea
where $\beta$  and $\gamma$  are constants, which are associated with the conserved quantities: the kinetic energy ($ \int d{\bf r} u^2/2 $) and kinetic helicity ($ \int d{\bf r} ({\bf u} \cdot \boldsymbol{\omega}) $) in 3D, and kinetic energy and enstrophy ($ \int d{\bf r} \omega^2/2 $) in 2D, where $ \boldsymbol{\omega} $ is the vorticity field. \citet{Onsagar:Nouvo1949_SH} modelled 2D Euler flow using a collection of point vortices that interact with logarithmic potential. Further, Onsagar  analyzed the above system and predicted  {\em negative temperature} and a large  cluster of same-circulation vortices for large energies.

There have been numerous efforts to verify the aforementioned predictions of Kraichnan~\cite{Kraichnan:JFM1973} and Lee~\cite{Lee:QAM1952}.   Most numerical works start with ordered initial condition that asymptotically approach the above equilibrium state. For example,  \citet{Cichowlas:PRL2005} simulated 3D Euler turbulence using a large-scale Taylor-Green vortex as an initial condition.  They observed Kolmogorov's spectrum for the intermediate wavenumbers and $k^2$ spectrum at large wavenumbers, which is a combination of nonequilibrium and equilibrium states. In later works, Krstulovic et al.~\cite{Krstulovic:PRE2009} simulated truncated Euler equation with a large-scale helical flow as an initial condition, and obtained Kraichnan's helical absolute equilibrium state at small scales.    Dallas et al.~\cite{Dallas:PRL2015} and  Alexakis and Brachet~\cite{Alexakis:JFM2019,Alexakis:JFM2020} studied Kolmogorov flow where the forcing is employed at intermediate scales.  They observed that the flow at scales larger than the forcing scale reaches a thermal equilibrium and exhibits $k^2$ energy spectrum. Similar behaviour has been observed for time-dependent projected Gross-Pitaevskii equation~\cite{Davis:PRL2001}, as well as  for truncated dissipation-less Burgers equation~\cite{Majda:PNAS2000,Ray:PRE2011}.  Also refer to earlier reviews by \citet{Orszag:CP1973,Kraichnan:ROPP1980}.

Recently, \citet{Verma:PTRSA2020}  employed $\delta$-correlated velocity field as initial condition and  performed preliminary simulation of 3D Euler turbulence.  They observed $k^2$ energy spectrum from the beginning itself. In the present paper, we perform detailed simulations  with  and without kinetic helicty and report equilibrium solutions for both 2D and 3D Euler equation, i.e., Eqs.~(\ref{eq:Euler2D}, \ref{eq:Euler3D}). In these solutions, the velocity field is as random as in thermodynamic gas.  In addition, we show that the kinetic energy flux is zero and that detailed balance in energy transfers is respected for the equilibrium solutions. Interestingly, similar solutions have been observed for dissipation-less Burgers equation and Korteweg–de Vries (KdV) equation when they were started with $\delta$-correlated fields~\cite{Verma:PRE2022}.

Euler turbulence exhibits interesting nonequilibrium behaviour that are different for 2D and 3D. The nonequilibrium  states are obtained for ordered initial condition, as in \citet{Cichowlas:PRL2005}. Customarily, 3D conservative systems thermalize when they are started with a nonequilibrium configuration.  As shown by \citet{Cichowlas:PRL2005},  3D Euler turbulence thermalizes in this spirit.  Regarding  2D Euler turbulence,  for coherent velocity field as an initial condition, $ E(k) $ differs significantly from Eq.~(\ref{eq:Euler2D}).  For example,  \citet{Fox:PF1973}  reported deviations from Eq.~(\ref{eq:Euler2D})   at  small wavenumbers for enstrophy-dominated 2D Euler turbulence.    \citet{Seyler:PF1975} observed  large vortex structures, similar to those in a discrete vortex system~\cite{Joyce:JPP1973}.  \citet{Verma:arxiv2022} showed that for several ordered initial condition,   2D Euler remains out of equilibrium throughout its evolution. These works indicate  2D Euler turbulence is out of equilibrium.    Even though, 2D Euler turbulence is in nonequilibrium state, \citet{Robert:JFM1991}, and \citet{Bouchet:PR2012} analyzed such structures in the framework of equilibrium statistical mechanics.

Euler equation is time reversible due to an absence of viscous dissipation. This is another reason why  thermodynamic entropy of Euler turbulence is constant. Note, however, that  the Euler equation exhibits irreversibility due to its chaotic nature.  The hydrodynamic entropy, to be describe in this paper, captures the irreversibility and disorder of Euler turbulence quite well. 

The structure of the paper is as follows. In Sections \ref{sec:eqns} we review the analytical works on equilibrium states of Euler turbulence. Section \ref{sec:entropy} summarizes the hydrodynamic entropy formalism, whereas Section \ref{sec:num_method} covers the numerical procedure for solving Euler equations. Sections \ref{sec:equilibrium_behaviour} covers the numerical results on equilibrium properties of Euler turbulence, whereas Sections \ref{sec:noneq_3D} and \ref{sec:noneq_2d} cover the nonequilibrium properties of 3D and 2D Euler turbulence respectively.  We conclude in Section \ref{sec:conclusions}.

\section{GOVERNING EQUATIONS AND EQUILIBRIUM STATES} 
\label{sec:eqns}
The incompressible Euler equation, which is the hydrodynamic equation with zero  external force and zero viscosity, is    
\bea
\partial_t \mathbf{u} + (\mathbf{u}\cdot \nabla) \mathbf{u} & = & -\nabla p, \label{eq:NS} \\
\nabla \cdot \mathbf{u} & = & 0, 
\eea
where  ${\bf u}$ and $p$ are the velocity and pressure fields respectively.  In Fourier space,  Eq.~(\ref{eq:NS}) gets transformed	to~\cite{Leslie:book,Frisch:book}
\be
\frac{d}{dt} \hat{\bf u}({\bf k}) + i  \sum_{\bf r} {\bf k} \cdot \hat{\bf u}({\bf q}) \hat{\bf u}({\bf r}) =   -i {\bf k} p({\bf k})
\label{eq:Fourier_space}
\ee
where ${\bf q = k-r}$, and $ \hat{\bf u}({\bf k}), p({\bf k})$ are the Fourier transforms of $ \mathbf{u}({\bf x}) $ and $ p({\bf x}) $ respectively.   
The \textit{modal energy} corresponding to $  {\bf \hat{u}(k)} $ is defined as  $ E({\bf k}) = | {\bf \hat{u}(k)}|^2/2 $, whose evolution equation is \cite{Kraichnan:JFM1959,Verma:book:ET}    
\be
\frac{d}{dt} E({\bf k})  =  \sum_{\bf p} \Im [ \{ {\bf k \cdot \hat{u}(q)  \}  \{ \hat{u}(r) \cdot \hat{u}^*(k)}\} ] = T({\bf k}),
\label{eq:E_k}
\ee
where $ \Im[.] $ represents the imaginary part of the argument, and $ T({\bf k}) $ represents the nonlinear energy transfer to Fourier mode $ {\bf \hat{u}(k)}$ from all other modes. Note that the pressure term of the Euler equation does not contribute to the evolution of $ E({\bf k}) $.

Under a statistical steady state, the average  rate of change of kinetic energy is zero, i.e., $ d\la E({\bf k}) \ra/dt =0 $.  Hence,  Eq.~(\ref{eq:E_k}) yields the following relation  for all $ {\bf k} $'s, 
\be
\la T({\bf k}) \ra = 0.  
\label{eq:zero_Tk}
\ee
The energy flux $ \Pi(k)  $ is defined as the net nonlinear energy transfer  from all the modes residing inside the sphere of radius $ k $ to the modes outside the sphere.  In terms of $ T({\bf k}) $, the energy flux $ \Pi(k)  $  is defined  as~\cite{Kraichnan:JFM1959,Verma:book:ET}
\be
\Pi(k) = -\sum_{|{\bf }k'|=0}^k T({\bf k'}).
\label{eq:Tk_Pik}
\ee
Therefore, under  statistical steady state,~\cite{Verma:PRE2022}  
\be
\la \Pi(k) \ra = 0.
\label{eq:zero_flux}
\ee
Thus, for a steady state, the energy flux for  Euler equation is zero.    The condition $ \la T({\bf k}) \ra = 0  $ for all $ {\bf k} $'s indicates that the system respects \textit{detailed balance in energy transfers}.  Thus, we can claim that the system is under equilibrium, as in thermodynamic equilibrium.   This general condition is expected to work for   conservative systems. Later in this paper we show that Eq.~(\ref{eq:zero_flux}) is satisfied for 3D Euler turbulence, but not for 2D Euler turbulence.

Liouville’s theorem is often invoked to characterize the statistical properties of a system in equilibrium.    For example, Gibbs measure is  an invariant measure of a system satisfying Liouville’s theorem~\cite{Huang:book:StatMech,Majda:PNAS2000}.  On this basis, Lee~\cite{Lee:QAM1952} and Kraichnan~\cite{Kraichnan:JFM1973}  derived that $ E(k) \sim k^2 $ for nonhelical Euler turbulence, and 
\be
E(k) \sim \frac{k^2}{\beta^2 - \gamma^2 k^2} 
\label{eq:Ek_Hk_Kraichnan}
\ee
 for helical turbulence, where $\gamma$ and $\beta$ are constants. Note, however, that Eqs.~(\ref{eq:E_k}, \ref{eq:zero_Tk}, \ref{eq:Tk_Pik}), which are related to the flux formalism, provide alternative framework for classifying equilibrium systems.

In the following discussion, we derive equilibrium energy distribution for 3D Euler turbulence using helical basis~\cite{Waleffe:PF1992,Verma:book:ET}.  The helical Fourier modes associated with wavenumber $ {\bf k} $ are~\cite{Waleffe:PF1992,Sagaut:book,Verma:book:ET}
\be 
u_+(\mathbf{k}), \quad u_-(\mathbf{k}).
\ee
We denote the corresponding energies as $ E_+(\bf k) $ and $ E_-(\bf k) $ respectively. Under equilibrium, $ E_+$ is distributed among all the positive helicity modes,  $ E_+(\bf k) $.  Similar distribution takes place for $ E_-({\bf k}) $.  The average values of $ E_+(\bf k) $ and $ E_-(\bf k) $ are $\la  E_+(\bf k) \ra$ and $\la  E_-(\bf k)  \ra$ respectively.  Note that
\bea
E({\bf k}) & = & E_+({\bf k}) + E_-({\bf k}), \label{eq:Ek_epm} \\
H({\bf k}) & = & k(E_+({\bf k}) - E_-({\bf k})),
\label{eq:Hk_epm} 
\eea
where $E({\bf k})$ and $H({\bf k})$ are modal energy and modal kinetic helicity respectively.
Hence,  $\la  E_+(\bf k) \ra \ne \la  E_-({\bf k})   \ra$ for helical flows.

For a given \textbf{k}, $ E_+(\bf k) $ and $ E_-(\bf k) $  fluctuate around their respective mean values. We denote the fluctuations in $ E_+(\bf k) $ and $ E_-(\bf k) $  using  $e_+$ and $e_-$ respectively. The probability for observance of $(e_+, e_-)$ is denoted by $P(e_+, e_-)$. Using conservation laws, we arrive at the following constraints:
\bea
\int_0^\infty P(e_+, e_-) de_+ de_-  & = & 1 \label{eq:prob_sum}\\
\int_0^\infty e_\pm P(e_+, e_-) de_+ de_-  & = &\la  E_\pm(\bf k) \ra, 
 \label{eq:prob_e_sum}
\eea
To obtain the distribution for the fluctuations $e_\pm$, we extremize the following function:
\bea
S & = &  -\sum P(e_+, e_-) \log(P(e_+, e_-)) \nonumber \\ 
&& -\alpha(\sum P(e_+, e_-) -1)  \nonumber \\
&& - \beta_+(\sum P(e_+, e_-) e_{+} -\la  E_+(\bf k) \ra)  \nonumber \\
&& -\beta_-(\sum P(e_+, e_-) e_{-} -\la  E_-(\bf k) \ra),
\eea
where $\sum$ is a shorthand for $\int_0^\infty  de_+ de_-$. Taking derivative of $ S  $ with relative to $ P(e_+, e_-) $ yields
\be
-\log(P(e_+, e_-)) -1 -\alpha - \beta_+ e_{+} - \beta_- e_{-}  = 0.
\ee
Hence, 
\be
P(e_+, e_-) = \exp[-(\alpha +1)] \exp(-\beta_+ e_{+}- \beta_- e_{-}).
\label{eq:P_inter}
\ee
Substitution of Eq.~(\ref{eq:P_inter}) in  Eq.~(\ref{eq:prob_sum}) yields
\be
\exp[-(\alpha +1)] = \beta_+ \beta_-.
\ee
Hence,
\be
P(e_+, e_-) = \beta_+ \beta_- \exp(-\beta_+ e_{+}- \beta_- e_{-}).
\label{eq:prob_equilibrium}
\ee
Substitution of the above in Eq.~(\ref{eq:prob_e_sum}) yields
\be
\la  E_\pm (\bf k)  \ra  = 1/\beta_\pm.
\label{eq:Epm_avg}
\ee 
Note that $\beta_\pm$ can be a function of $k$, and that
\be
e_\pm = \frac{1}{2} (e \pm h/k),
\ee
where $e$ and $h$ are the kinetic energy and kinetic helicity respectively. Substitution of the above expressions in Eq.~(\ref{eq:prob_equilibrium}) yields
\be
P(e, h) = (\beta^2 - \gamma^2 k^2) \exp(-\beta e - \gamma h),
\label{eq:prob_e_h}
\ee
where
\be
\beta_\pm = \beta \pm \gamma k .
\label{eq:beta_pm}
\ee
Using Eqs.~(\ref{eq:Ek_epm}, \ref{eq:Hk_epm}, \ref{eq:Epm_avg},  \ref{eq:beta_pm}) we derive 
\bea
\la  E(\bf k) \ra & = & \frac{2\beta}{(\beta^2 - \gamma^2 k^2)},  \label{eq:Ek} \\
\la  H(\bf k) \ra & = & -\frac{2 \gamma k^2}{(\beta^2 - \gamma^2 k^2)} . \label{eq:Hk}
\eea
 
A nonhelical flow is a special case with $ \gamma =0 $ or $ \beta_+ = \beta_- $.  For this case, $ \la  E(\bf k) \ra = 2/\beta $ and $\la  H(\bf k) \ra  =0  $, and the velocity components in Craya-Herring basis are independently random.  However, for helical case, correlations develop among the Fourier modes.  As is evident from  Eq.~(\ref{eq:Epm_avg}, \ref{eq:beta_pm}, \ref{eq:Ek}), the total kinetic energy $ E $, as well as $E_\pm$, are not equipartitioned among all the Fourier modes.  Also note that the shell spectra for the energy and kinetic helicity are
\bea
\la E(k) \ra = 4\pi k^2 \la  E({\bf k}) \ra, \\
\la H(k) \ra = 4\pi k^2 \la H({\bf k}) \ra.
\eea
In our future discussion, we will drop the angular brackets for $E(k)$ and $E_\pm(k)$. Also note that for nonhelical 3D Euler turbulence,
\be
E(k) = \frac{8 \pi k^2}{\beta}.
\label{eq:Ek_nonhelical_formula}
\ee
We will compare the above expression with our numerical data.

The equilibrium energy spectrum for 2D Euler turbulence has been derived following similar lines as above~\cite{Kraichnan:JFM1973,Lesieur:book:Turbulence}. Using the fact that the energy and enstropy are conserved for 2D Euler turbulence, one can derive that 
\be
P(e) = (\beta + \gamma k^2) \exp[-(\beta + \gamma k^2) e],
\label{eq:prob_e_2D}
\ee
and
\be
E({\bf k}) = \frac{1}{\beta + \gamma k^2}.
\ee
Hence, the shell spectrum for the energy for 2D Euler turbulence is $E(k) = 2\pi k E({\bf k}) = 2\pi k/(\beta+\gamma k^2)$.

The  connection mentioned above between the $\delta$-correlated velocity field (white noise) and $k^2$ spectrum provides a hint that we should choose white noise as the initial condition for the equilibrium solution of Euler turbulence. We follow this strategy in the present paper.


\section{Hydrodynamic Entropy}
\label{sec:entropy}

The thermodynamic  entropy of Euler turbulence is  constant due to an absence of viscosity~\cite{Landau:book:Fluid}. However,  the disorder in Euler turbulence varies with time for nonequilibrium scenario. Hence, the thermodynamic entropy is not suitable for  quantifying order in Euler turbulence.  Therefore, \citet{Verma:arxiv2022} defined ``hydrodynamic entropy" using Shannon's formula~\cite{Shannon:BELL1948}. They postulated that  the probability of occurrence of a Fourier mode with wavenumber $ {\bf k} $ is $ p_{\bf k} = E({\bf k})/E$, where $ E({\bf k}) $ is the modal energy, and $ E $ is the total energy. In terms of these quantities, the hydrodynamic entropy of the flow was defined as \cite{Verma:arxiv2022}
\be
S = -\sum_{\bf k}   p_{\bf k}  \log_2 ( p_{\bf k} ).
\label{eq:entropy}
\ee 
The above entropy can be used to quantify hydrodynamic order of a snapshot of  a fluid flow. Note that  hydrodynamic entropy differs from  the thermodynamic entropy, which depends on temperature and volume of the system.  

\section{Numerical Details} 
\label{sec:num_method}

We perform pseudo-spectral simulation~\cite{Boyd:book:Spectral,Canuto:book:SpectralFluid} of Euler flow using our code TARANG~\cite{Verma:Pramana2013tarang,Chatterjee:JPDC2018}. We simulate both equilibrium and nonequilibrium configurations of Euler turbulence.  A key issue in Euler turbulence simulation is energy conservation, which is not conserved while using standard time stepping schemes, e.g., Runge-Kutta method. In this paper, we employ {position-extended Forest-Ruth-like} (PEFRL) scheme~\cite{Omelyan:CPC2002,Forest:PD1990} (see Appendix~\ref{sec:appendix})  that conserves energy to high precision.  We also dealise the code using {\em two-third rule}~\cite{Boyd:book:Spectral,Canuto:book:SpectralFluid}.

For studying nonequililbrium behaviour of Euler turbulence, we employ large-scale flow structures as initial condition. For example, Taylor-Green vortices are employed for 3D Euler equation~\cite{Cichowlas:PRL2005}. However, we employ $\delta$-correlated velocity field as initial condition to study equilibrium properties of Euler turbulence. 



We implement the above random initial condition using Craya-Herring basis~\cite{Craya:thesis,Herring:PF1974,Sagaut:book,Verma:book:ET} whose unit vectors for a wavenumber ${\bf k}$ are 
\bea
 \hat{e}_1(\textbf{k}) & = & ( \hat{k} \times \hat{n})/|\hat{k} \times \hat{n}| , \\
 \hat{e}_2(\textbf{k}) & = &  \hat{k} \times \hat{e}_1(\textbf{k}) ,
 \eea
  where $\hat{n}$ is chosen as any direction, and $ \hat{k} $ is the unit vector along \textbf{k}.  In this basis, the 3D incompressible velocity field is 
  \be
  {\bf u}(\textbf{k}) = u_1(\textbf{k}) \hat{e}_1 (\textbf{k}) + u_2 (\textbf{k})  \hat{e}_2(\textbf{k}),
  \ee
   while 2D incompressible velocity field is
   \be
   {\bf u}(\textbf{k})  =  u_1(\textbf{k})  \hat{e}_1(\textbf{k}).
   \ee  

For simulating 3D nonhelical flows (zero kinetic helicity),  we start with $u_1({\bf k})=0$ and $u_2({\bf k}) = \sqrt{2E/M} \exp(i \phi_2({\bf k}))$, where $E$ is the total kinetic energy, $M$ is the total number of dealiased modes, and the phase $\phi_2({\bf k})$ is chosen to be a random number from uniform distribution in a band of $[0,2\pi]$.  For the 2D simulation, we take $u_1({\bf k}) = \sqrt{2E/M} \exp(i \phi_1({\bf k}))$ with random phase for $\phi_1({\bf k})$.  For simulating helical Euler turbulence with kinetic helicity $H$, we choose $|u_1({\bf k})| = |u_2({\bf k})| = \sqrt{E/M} $ and  random $\phi_2({\bf k})$ from a uniform distribution in $[0,2\pi]$.  The other phase is $\phi_1({\bf k}) = \phi_2({\bf k}) - \sin^{-1}\sigma_c({\bf k})$, where  $\sigma_c({\bf k}) =   \Re[ {\bf u^*({\bf k})} \cdot \boldsymbol{\omega}({\bf k)}]/ (k |\textbf{u}(\textbf{k})|^2) $.  We use $\sigma_c({\bf k})$ to control $H$~\cite{Sadhukhan:PRF2019}. Note that the total number of active dealiased modes $M = (2N/3)^3$ for 3D and $M = (2N/3)^2$  for 2D, where $N$ is the number of grid points in each direction.

\section{EQUILIBRIUM Behaviour of Euler turbulence} 
\label{sec:equilibrium_behaviour}
In this section, we report the equilibrium nature of 2D and 3D Euler turbulence.  We discuss the randomness of the flows using real space density plots of vorticity,  the probability distribution functions, as well as energy spectra and fluxes. 

\subsection{For 3D Euler equation}
\label{subsec:KE}

We perform spectral simulation using $\delta$-correlated velocity field as an initial condition. See Sec.~\ref{sec:num_method} for details. We employ $128^3$ grid and PEFRL scheme for time stepping with $dt = 10^{-4}$. We perform two simulations, one without kinetic helicity, and the other with kinetic helicity. The total kinetic energy is 0.019 for the nonhelical (without kinetic helicity) flow. For the helical flow, the total kinetic energy and kinetic helicity are 0.038 and 2.8 respectively.  We perform these simulations up to 25 and 30 eddy turnover time units respectively.   

For the nonhelical Euler turbulence,  the total kinetic energy is conserved to 13 significant digits ($E = 0.019 \pm 9 \times 10^{-13}$), while the kinetic helicity remains negligible  throughout (rms value of $6\times 10^{-12}$).  For the helical turbulence, we choose $\sigma_c(\textbf{k} )= 0.9$ for all \textbf{k}'s to inject significant kinetic helicity.    Here, the total kinetic  energy and total kinetic helicity for the run are $0.038 \pm 7 \times 10^{-12} $ and $ 2.8 \pm 7 \times 10^{-9}  $ respectively; thus, they remain conserved throughout the run.

\begin{figure}
	\includegraphics[width=8.5cm]{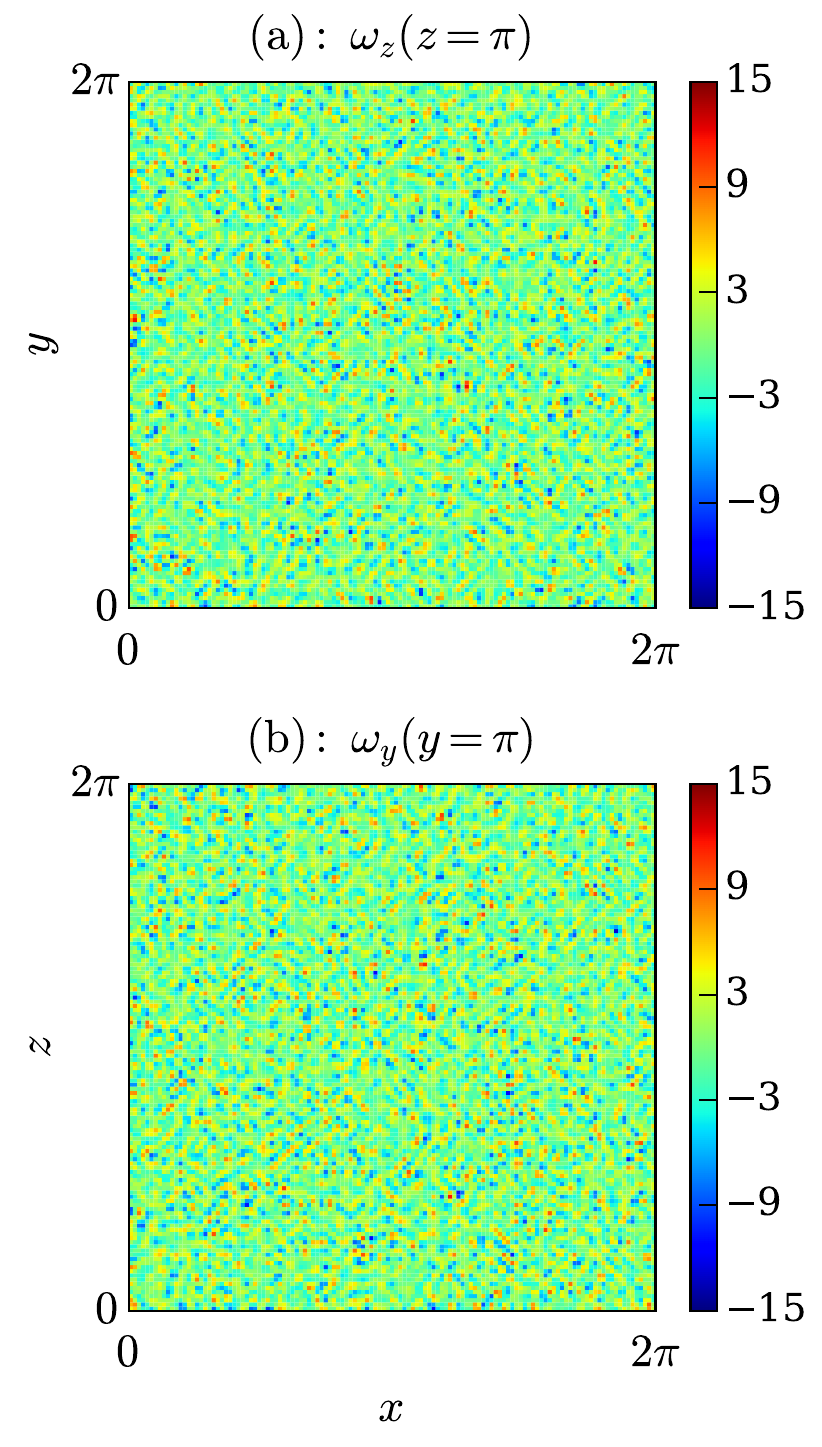} 
	\caption{(Color online)  For 3D nonhelical flow with $\delta$-correlated velocity as initial condition: Density plots of the perpendicular component of vorticity on (a) horizontal midplane ($z= \pi$), and (b) vertical midplane ($y= \pi$).  These plots indicate random velocity configuration.
	}\label{fig:Vorticity_plot_3d}
\end{figure}

Now we describe the equilibrium properties of Euler turbulence using the results of the above runs. We observe that  the nonhelical flow remains random, as in white noise, at all times.  Note, however, that the amplitudes and phases of all the modes vary randomly with time.  In Fig.~\ref{fig:Vorticity_plot_3d}(a,b), we exhibit the density plots of the perpendicular vorticity components of the flows in the horizontal and vertical mid planes for a snapshot. The plots clearly demonstrate the random nature of flow. 

\begin{figure}
	\includegraphics[width=8.5cm]{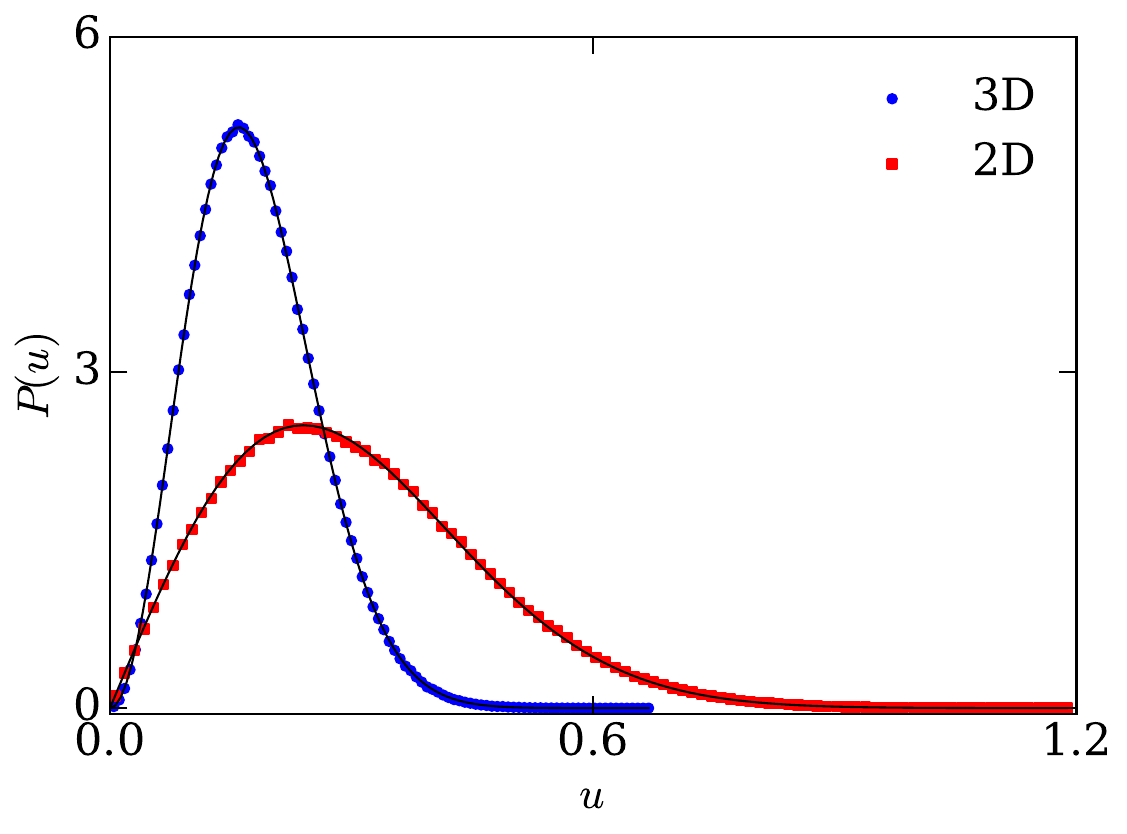} 
	\caption{(Color online) For the nonhelical Euler simulations: Probability distribution functions ($P(u)$) of the velocity magnitude for 2D (red
		squares) and 3D (blue circles) real-space flows. The numerical
		PDFs match closely with Maxwell-Boltzmann distribution for
		3D  and for 2D (black curves) thermodynamic
		systems.}
		\label{fig:PDFs}
\end{figure}

We substantiate the randomness of the flows   by computing the  probability distribution function (PDFs) of the magnitude of the real-space velocity field ($u$) of a snapshot, and test whether it obeys Maxwell-Boltzmann distribution, which for a 3D flow is 
\begin{equation}
P(u) = \sqrt{2/\pi}~a^{-3} u^2 \exp(-u^2/2a^2), 
\label{eq:Pu_3d}
\end{equation}
where   $a$ is the scale parameter. The numerical $P(u)$ for nonhelical 3D run, which is exhibited in Fig.~\ref{fig:PDFs},
matches quite accurately with  Eq.~(\ref{eq:Pu_3d}) with $a=0.11$. Hence, we claim that the velocity field of Euler turbulence is as random as the velocity distribution of  gas molecules in thermodynamics.

\begin{figure}
	\includegraphics[width=8.5cm]{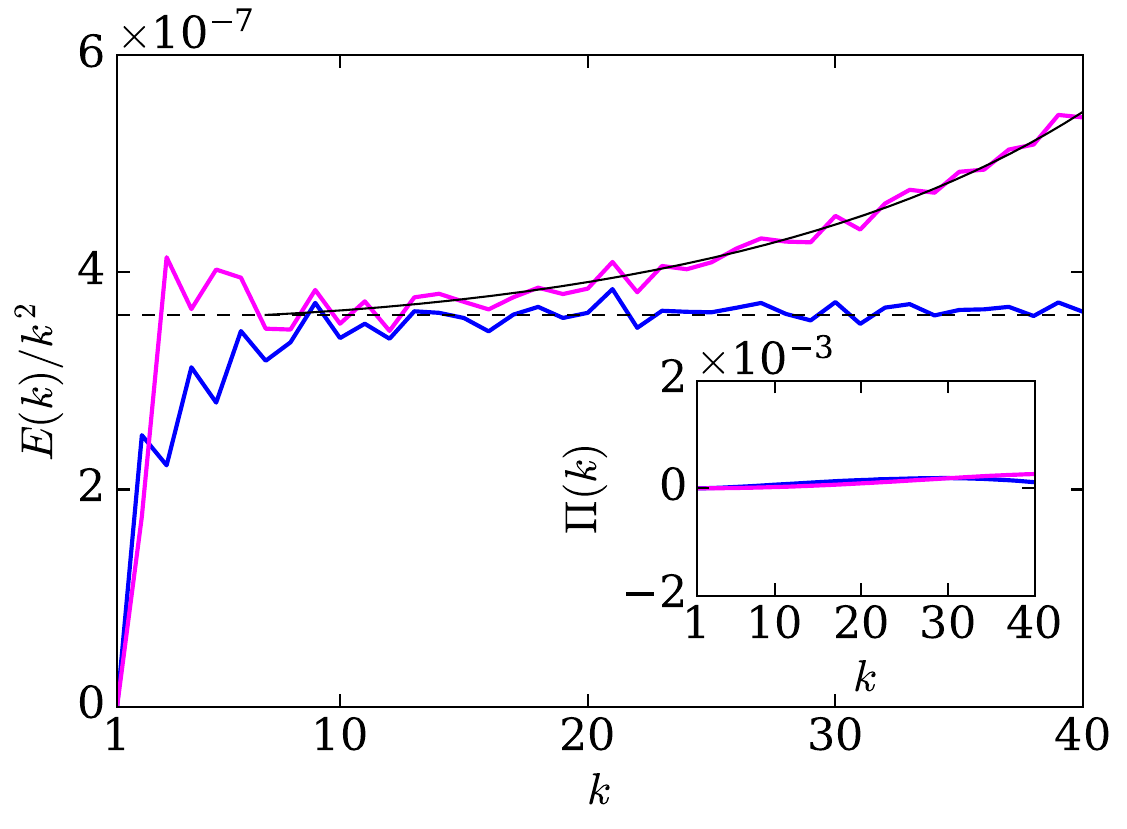} 
	\caption{(Color online) Plot of normalized energy spectra $E(k)/(k^2)$ for the  nonhelical simulation [blue curve] and the helical run [magenta curve].  The numerical plots match reasonably well with Kraichnan's predictions [Eq.~(\ref{eq:Ek})], which are shown as dashed lines. The inset shows the energy fluxes for the these runs using the same colour. 
	}\label{fig:Spectrum_3D}
\end{figure}

Next, we compute the energy spectrum and flux for the nonhelical simulation.  As shown in Fig.~\ref{fig:Spectrum_3D}, the energy flux is zero (apart from fluctuations).  Besides, the normalized energy spectrum, $E(k)/k^2$, is flat around $8 \pi/\beta$ with $\beta =0.64\times 10^{8} $ [see Eq.~(\ref{eq:Ek_nonhelical_formula})]. Hence, we claim that  $ E(k) $ varies as $k^2$ for the whole range of wavenumbers.   These results are consistent with the $\delta$-correlated (white noise) nature of the real-space velocity field, thus validating the predictions of absolute equilibrium theory~\cite{Kraichnan:JFM1973,Lee:QAM1952}. The vanishing energy flux follows from the fact that $ \delta $-correlated velocity field has zero triple correlation.   In addition, using field-theoretic arguments, Verma~\cite{Verma:PR2004,Verma:book:ET,Verma:JPA2022} has shown that the equipartitioned Fourier modes yield zero kinetic energy flux.  

The probability $P(e,h)$ of 3D Euler turbulence is given by Eq.~(\ref{eq:prob_e_h}). Since $H=0$ for the nonhelical run, we set $\gamma=0$ in the equation and compute $P(e)$. Unfortunately, the numerical data is not sufficient for the computation of $P(e)$  for a given $E({\bf k})$, hence we sample all $E({\bf k})$'s of a snapshot to compute $P(e)$. Figure~\ref{fig:P(E,H)} exhibits thus computed $P(e)$. We observe that the numerical data is described by Eq.~(\ref{eq:prob_e_h}) quite well with $\beta = 0.64\times 10^8$, which is consistent with the energy spectrum computation.
\begin{figure}
	\includegraphics[width=8.5cm]{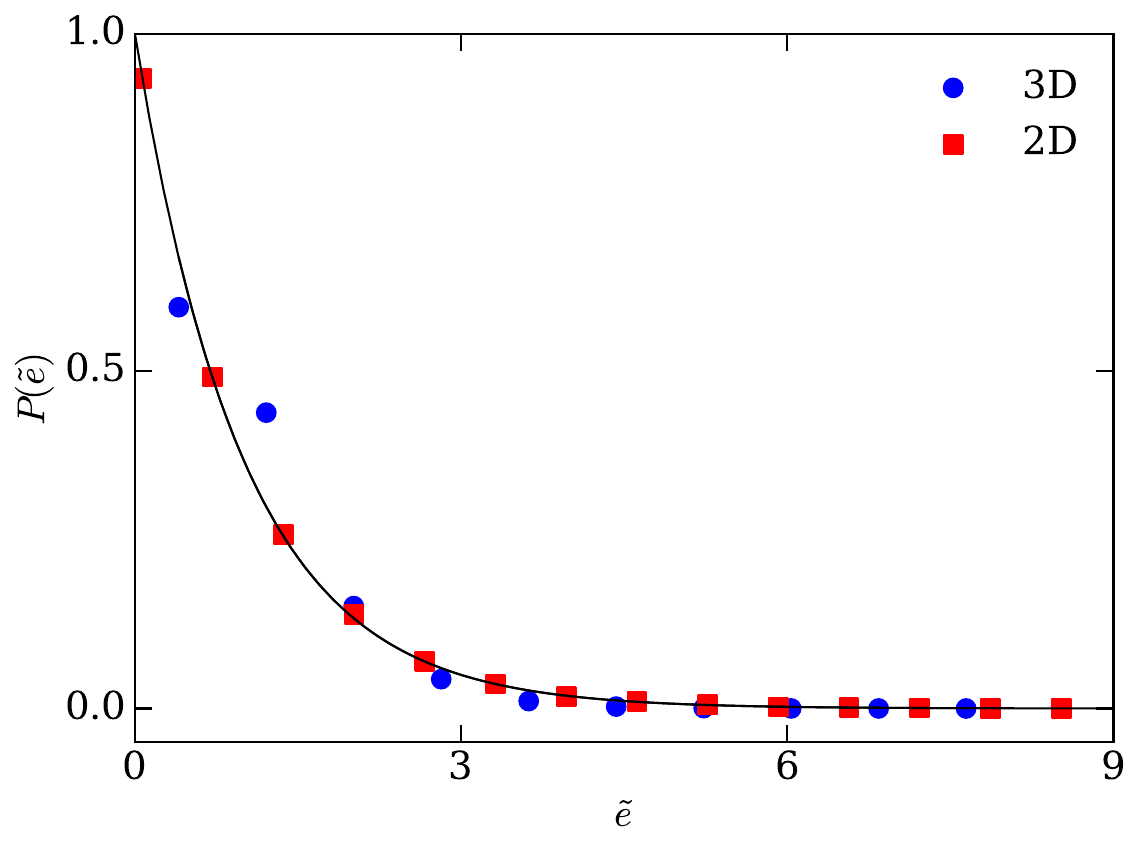} 
	\caption{(Color online) For nonhelical 3D Euler and 2D Euler runs, plots of $P(e)$  at $t = 20$. Here, $\tilde{e}=\beta e$ with  $\beta  = 0.64\times 10^{8}$ for 3D and 	$0.8\times 10^{7}$ for  2D  simulation.}
	\label{fig:P(E,H)}
\end{figure}

\begin{figure}
	\includegraphics[width=8.5cm]{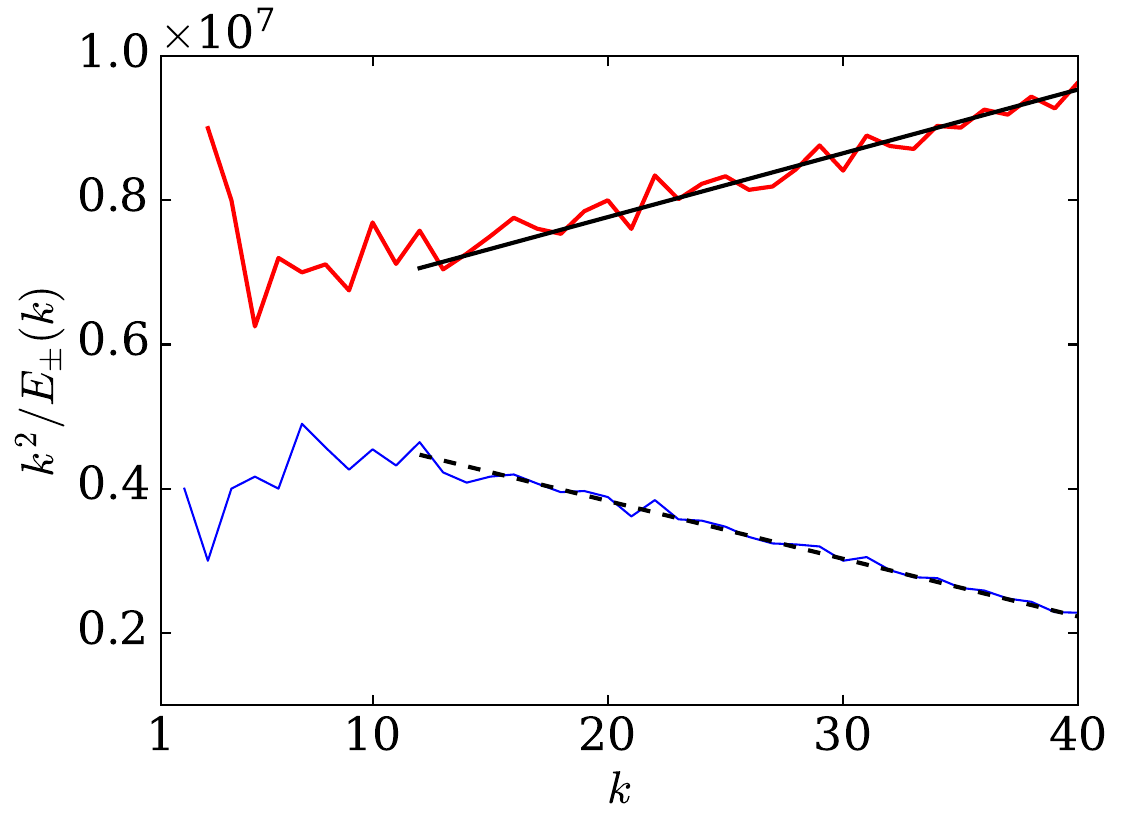} 
	\caption{(Color online) For 3D helical simulation, plots of  $k^{2}/E_{+}(k)$ (blue curve) and $k^{2}/E_{-}(k)$ (red curve) for the helical simulation. The best fit curves, shown as black dashed and solid lines, are consistent with Eq.~(\ref{eq:Epm_avg}).  } 
	\label{fig:Epm(k)}
\end{figure}

Now, we report the energy spectra, $ E_\pm(k), E(k) $, and energy flux for the helical run. We observe that $E_\pm(k)$ are consistent with Eqs.~(\ref{eq:Epm_avg},\ref{eq:beta_pm}) with $ \beta = 0.7\times10^8 $ and $\gamma = 1.1\times 10^6$. We illustrate $E_\pm(k)$ in Fig.~\ref{fig:Epm(k)}.  We also compute $ E(k) $ and plot it in  Fig.~\ref{fig:Spectrum_3D}. We find that $ E(k) $  matches with Eq.~(\ref{eq:Ek}) quite well with $ \beta,\gamma $ computed above using $ E_\pm(k) $.  The deviation of $E(k)$  from $k^2$ at large $k$ is consistent with Eq.~(\ref{eq:Ek}).  Also, as shown in Fig.~\ref{fig:Spectrum_3D}, the energy flux vanishes for the helical 3D Euler turbulence as well. Based on the above results, it is evident  that the velocity field of the helical run too is under equilibrium.  For small wavenumbers, a small deviation  of $E(k)$ from Eq.~(\ref{eq:Ek})  may be due to an asymmetry in the energy transfers. For example, $ \textbf{u}(\textbf{k}_0  =1) $ has no Fourier mode with wavenumber less than $\textbf{k}_0$.

\subsection{For 2D Euler equation}
We simulate  2D Euler turbulence on a $1024^2$ grid using the method described in Sec.~\ref{sec:num_method}.  We carry out the 2D simulation up to 30 time units with $ dt = 10^{-3}$. The total energy and total enstrophy of the flow are conserved    with good accuracy, for example, $E = (0.058 \pm 9 \times 10^{-9})$ and $E_{\Omega} = 4534 \pm 10^{-3}$.

\begin{figure}
\includegraphics[width=8.5cm]{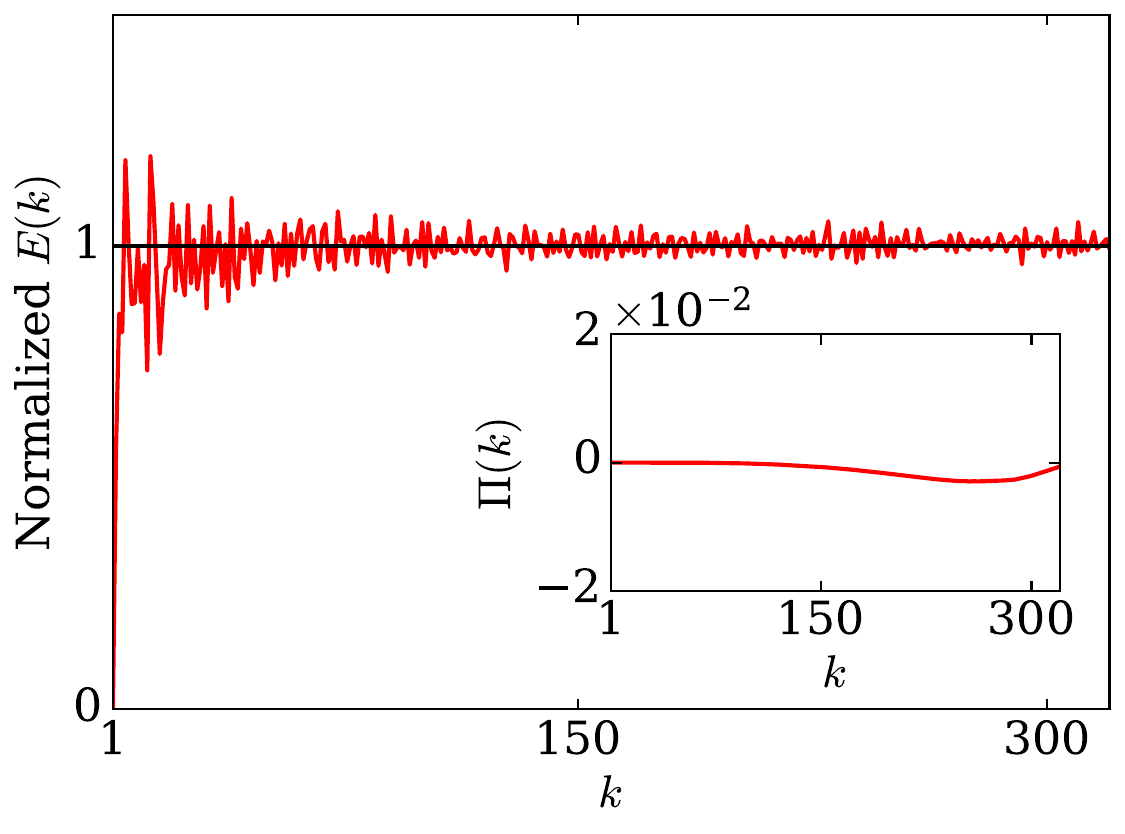} 
\caption{(Color online) For 2D Euler simulation, plot of the normalized energy spectrum $ \beta E(k)/(2 \pi k)$ with $\beta = 0.8\times 10^7$. The energy flux is in the inset.}
\label{fig:Ek_2D}
\end{figure}

For 2D Euler run too, the flow is random as a thermodynamic gas. In Figs.~\ref{fig:PDFs} and \ref{fig:P(E,H)}, we illustrate the PDF of $u$ and $e$, both of which show equilibrium properties similar to 3D Euler turbulence.  Note that the Maxwell-Boltzmann distribution for a 2D flow is 
\begin{equation}
	P(u) = a^{-2} u \exp(-u^2/2a^2), 
	\label{eq:Pu_2d}
\end{equation}
where   $a = 0.24$.  For our simulation with $ \delta $-correlated \textbf{u}, $\gamma \approx 0$ because $k_\Omega$, the centroid for enstrophy, exceeds the grid size. Hence, $E(k) = 2\pi k/\beta$, as is evident from Fig.~\ref{fig:Ek_2D}.  We obtain $\beta = 0.8 \times 10^7$, which is consistent with $P(e)$ plot of Fig.~\ref{fig:PDFs}. In addition, the energy flux is zero for this case, indicating detailed balance of energy transfer. 

We also study the phase space projection of the phase space trajectory on  the 
$ \{\Re(u(\mathbf{q})), \Im(u(\mathbf{q}))\}$ plane for  wavenumber $ \mathbf{q} = (128, 128) $. We observe a random scatter of the trajectory whose extent is around two times  $\sqrt{\la |\hat{u}(k)|^{2} \ra} = 0.5 \times 10^{-3}$.  We believe that the trajectory will wander off to a larger distance from the origin if we wait for a longer time.  These observations clearly indicate that 2D Euler turbulence exhibits equilibrium behaviour. 
\begin{figure}
	\includegraphics[width=8.5cm]{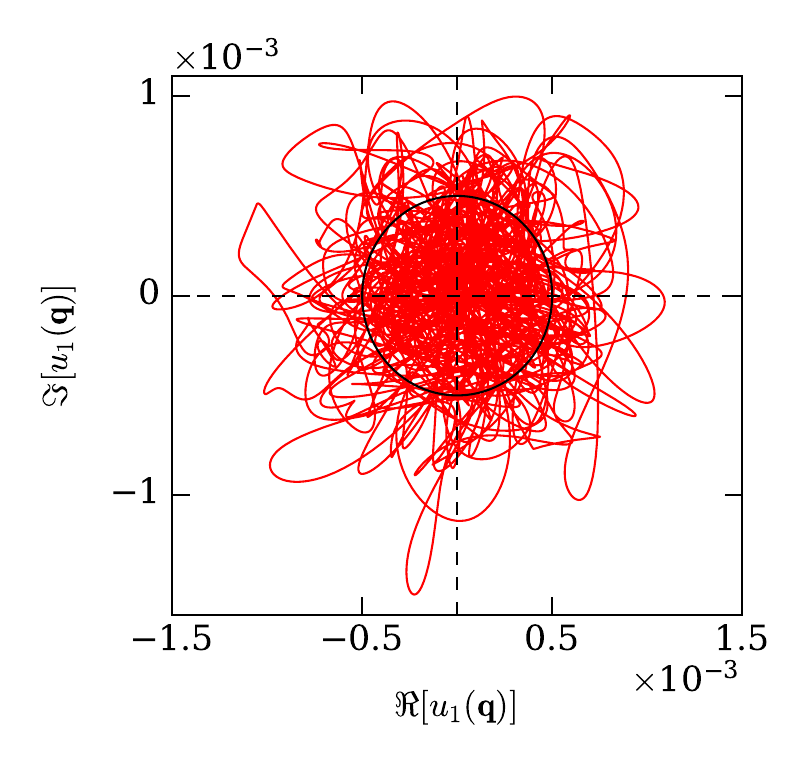} 
	\caption{(Color online) For 2D Euler simulation, phase space projection of phase space trajectory on the 
		$ \{\Re(u(\mathbf{q})), \Im(u(\mathbf{q}))\}$ plane for $ \mathbf{q} = (128, 128) $. The  radius of the black circle in the figure is $\sqrt{\la |\hat{u}(k)|^{2} \ra} = 0.5 \times 10^{-3}$. 
	}\label{fig5}
\end{figure}

In the next two sections, we describe nonequilibrium behaviour of 3D and 2D Euler turbulence. 

\section{Nonequilibrium behaviour and thermalization of 3D Euler turbulence} 
\label{sec:noneq_3D}

Cichowlas et al.~\cite{Cichowlas:PRL2005} reported thermalization  for 3D Euler turbulence, which is an important topic of research in nonequilibrium statistical mechanics, both classical and quantum.  They presented a model of thermalization that relates the energy of the thermalized modes to the transition wavenumber between the nonequilibrium and thermal modes.  Krstulovic and Brachet   \cite{Krstulovic:PD2008} constructed a two-fluid model of truncated Euler equation and determined the effective viscosity and thermal diffusion. They employed EDQNM closure and Monte-Carlo scheme for their derivation. Note that the intermediate stage of 3D Euler turbulence is in a mix state of equilibrium (large~$k$) and nonequilibrium (intermediate $k$). The flow thermalizes after tens of eddy turnover times.

\begin{figure}[hbtp]
	\includegraphics[scale=0.9]{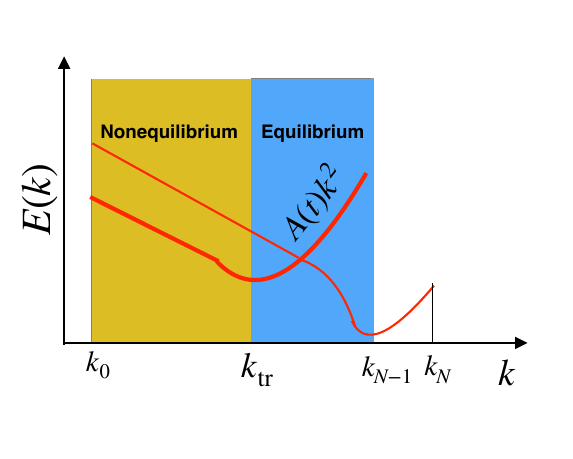}
	\caption{A schematic diagram exhibiting the evolution of energy spectrum $E(k)$ of 3D Euler turbulence during thermalization.  The thin red curve represents $E(k)$ during the early phase, while the thick red curve at an intermediate state.  At the transition wavenumber  $k_\mathrm{tr}$, $E(k)$ changes from $k^{-5/3}$ to $k^2$.  The two regimes, nonequilibrium and equilibrium, are represented by yellow and blue colors respectively.   }
	\label{fig:T_thermalization}
\end{figure}

In the following discussion, we estimate the time required for thermalization  in 3D Euler turbulence~\cite{Cichowlas:PRL2005}.  We denote the wavenumber shells in 3D Euler turbulence as $k_0, k_1, ...,k_{N-1}, k_N$, and assume that the  initial condition is a large-scale vortex    with wavenumber $k_0$, as in  \cite{Cichowlas:PRL2005}.  Nonlinear interactions transfer energy from $k_0$ to $k_1$, from $k_1$ to $k_2$, ..., $k_{N-1}$ to $k_N$.  The cascade however stops at $k = k_N$ where the energy piles up.  After sufficiently large  accumulation of energy at $k_N$, the energy starts to grow at  wavenumbers shell   $k_{N-1}$, and then at $k_{N-2}$,  and so on.  This is how the  large wavenumber shells acquire $k^2$ spectrum, as reported by \citet{Cichowlas:PRL2005} and \citet{Krstulovic:PD2008}. See Fig.~\ref{fig:T_thermalization} for an illustration.

Following Kolmogorov's theory of turbulence, the energy cascade rate to the large-wavenumber modes can be estimated as $\epsilon_u = U^3/L \sim U^3 k_0$, where $ L, U$ are the large-scale length and velocity respectively~\cite{Kolmogorov:DANS1941Dissipation,Kolmogorov:DANS1941Structure,Frisch:book,Lesieur:book:Turbulence}.  This energy flux accumulates at large wavenumbers and builds up $ A(t) k^2$ spectrum from the transition wavenumber $k_\mathrm{tr}$ to $k_\mathrm{max}$ (see Fig.~\ref{fig:T_thermalization}).  Therefore, in time $t$, 
\be
\epsilon_u t  \sim \int_{k_\mathrm{tr}}^{k_\mathrm{max}} A(t) k^2 dk,
\ee
or
\be
U^3 k_0 t \sim A(t) [k_\mathrm{max}^3 - (k_\mathrm{tr}(t))^3 ].
\label{eq:energetics}
\ee
Over time, $A(t)$ increases and $k_\mathrm{tr}(t)$ decreases. Using Eq.~(\ref{eq:energetics}) we can deduce the total  time taken for thermalization ($T$) as follows.  During the final stage, $k_\mathrm{tr} \rightarrow k_0 \ll k_\mathrm{max}$ and $A(T) \sim E/N^3$.  Hence,
\be
T \sim \frac{E}{N^3 U^3 k_0} k_\mathrm{max}^3 \sim \frac{L}{U}
\ee
because $k_\mathrm{max} \approx N/2$.  Thus, a 3D Euler flow with large-scale vortex as an initial condition is expected to thermalize in order of one eddy turnover time~\cite{Cichowlas:PRL2005,Krstulovic:PRL2011}.     In practice, this process takes tens of eddy turnover time~\cite{Cichowlas:PRL2005}. 

In Euler turbulence with coherent large-scale structures, the  energy in the inertial and dissipation range is converted incoherent (random) energy at small scales~\cite{Krstulovic:PD2008,Verma:EPJB2019}.   In the language of statistical mechanics, the yellow and blue regions of Fig.~\ref{fig:T_thermalization} could represent \textit{system} and \textit{heat bath} respectively. In this process, coherent energy is being converted to incoherent energy, thus giving a semblence of frictional effect~\cite{Krstulovic:PD2008,Krstulovic:PRE2009}. \citet{Verma:EPJB2019} argued that dissipation can emerge in conservative systems in a similar manner. Thus, the nonequilibrium and equilibrium states of Euler turbulence yield valuable  insights into the thermalization process in conservative systems.  

\begin{figure}
	\includegraphics[width=8.5cm]{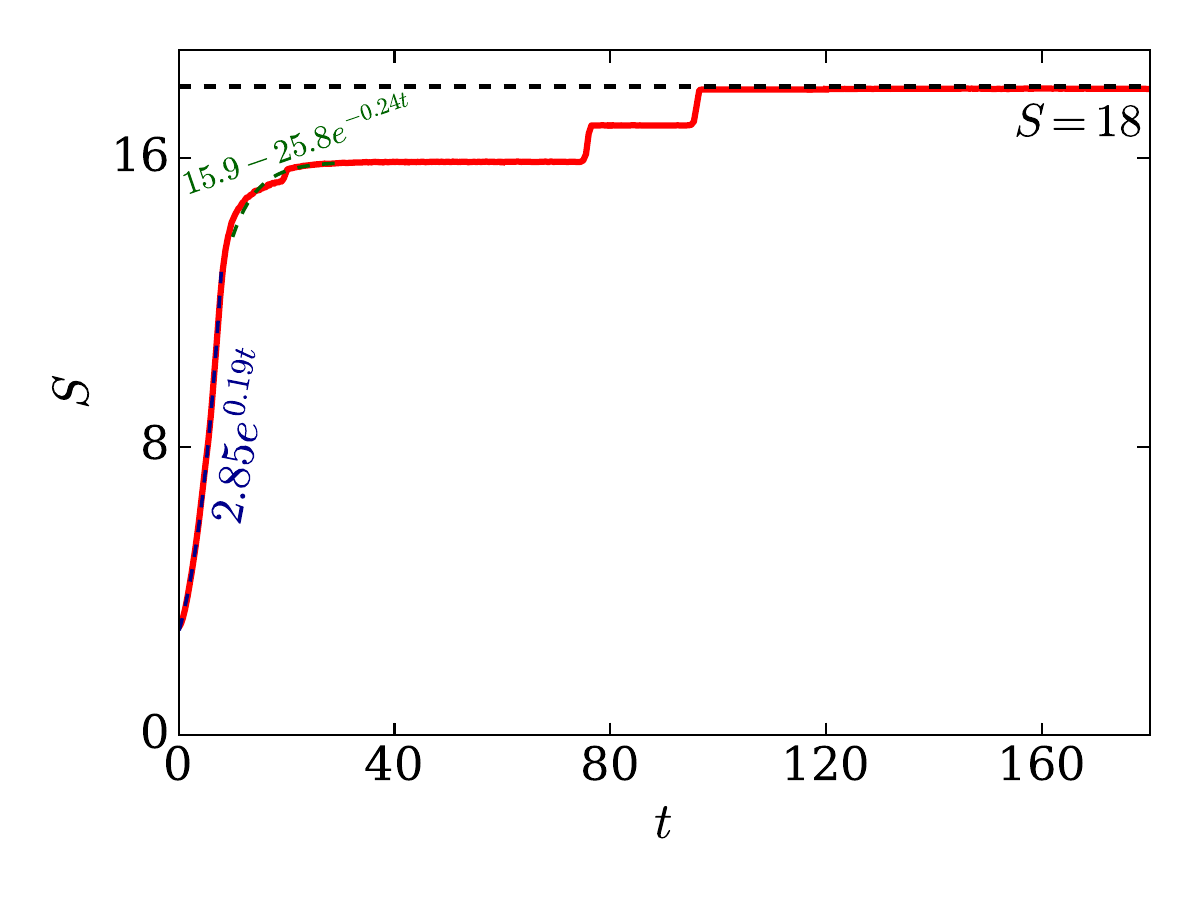} 
	\caption{(Color online) For 3D Euler turbulence, hydrodynamic entropy exhibits monotonic growth, with  exponential increase in the beginning, and saturation in the end. 
	}\label{fig:entropy_3d_3d_nq}
\end{figure}

Before we end this section, we describe the hydrodynamic entropy of 3D Euler turbulence, as reported by~\citet{Verma:arxiv2022}.  They simulated 3D Euler turbulence on a $ (2\pi)^3 $ box with a $ N^3 $ grid, where $ N=128$, using Taylor-Green vortex ($ k_0=1 $) as an initial condition. The  simulation was run up to 180 nondimensional time units using   $dt = 10^{-4} $ (see Sec.~\ref{sec:num_method}).  The total energy for the run is $0.125$ and it is conserved up to 12 decimal places.   The time evolution of hydrodynamic entropy computed using Eq.~(\ref{eq:entropy}) is shown in  Fig.~\ref{fig:entropy_3d_3d_nq}. We observe that the hydrodynamic entropy increases monotonically and asymptotes to 18, which is near the maximum possible entropy of 18.3.


\section{Nonequilibrium behaviour of 2D Euler turbulence}
\label{sec:noneq_2d}

In a recent work, \citet{Verma:arxiv2022} simulated 2D Euler turbulence with ordered initial condition and observed nonequilibrium behaviour. In this section, we briefly review their results. 

\begin{figure}
	\includegraphics[width=8.5cm]{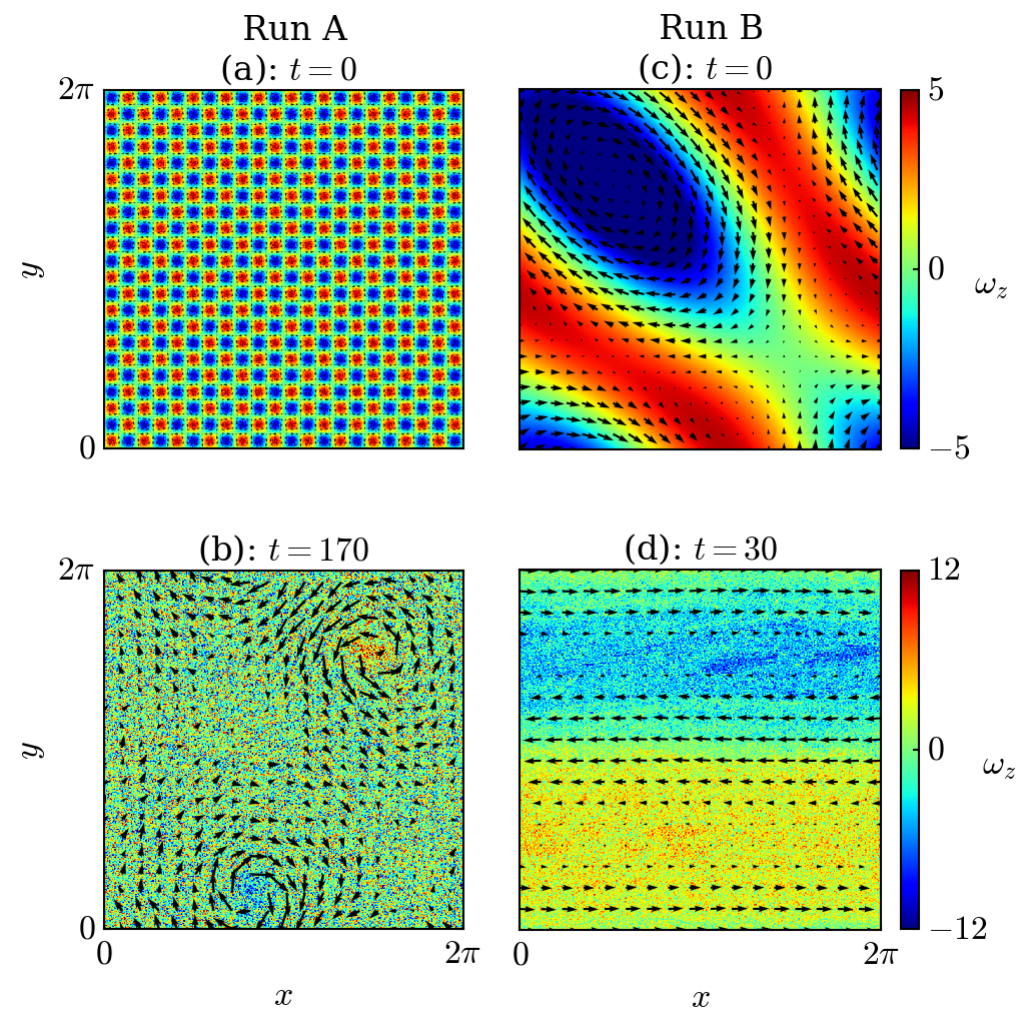} 
	\caption{(Color online) For Runs A and B of 2D Euler turbulence: (a,c) the initial states, (b,d) the final states respectively. Here we plot the velocity field over the density plots of the vorticity field.
	}\label{fig:2D_profile}
\end{figure}

\citet{Verma:arxiv2022} simulated 2D Euler turbulence on a $ (2\pi)^2 $ box using $ 512^2 $ grid.  See Sec.~\ref{sec:num_method} for  details. Two of \citet{Verma:arxiv2022}'s runs are given below:
\begin{enumerate}
	\item Run A: The initial velocity profile is taken as $ ( \sin 11x \cos 11y + \eta_x, - \cos 11x \sin 11y + \eta_y )  $, where $ (\eta_x, \eta_y) $ is random noise. We take $|\eta_x| \ll 1  $ and $|\eta_y| \ll 1$.
	
	\item Run B: The initial nonzero velocity Fourier modes are  $ {\bf u}(1,0) = (0,1)$,  $ {\bf u}(0,1) = (1,0)$, and  $ {\bf u}(1,1) = (-i,i)$.  
\end{enumerate}
These runs were time advanced up to 170 and 30  turnover times  ($ 2\pi/U_\mathrm{rms} $) respectively.  The Runs A and B  reach  steady states after 100 to 10 eddy turnover times respectively.  For Runs A and B, the total energy and enstrophy are (0.2500954, 4) and (62.17, 6) respectively, and they are conserved to many significant digits. The initial  and final  states of the two runs are shown on the top and bottom panels of Fig.~\ref{fig:2D_profile}.  Here, the velocity field is superposed over the density plots of the vorticity field.  The aysmptotic states of Runs A and B  are a vortex-antivortex pair~\cite{Onsagar:Nouvo1949_SH} and  a unidirectional flow (shear layer) respectively, which are embedded in small-scale noisy flow.  

\begin{figure}
	\includegraphics[width=8.5cm]{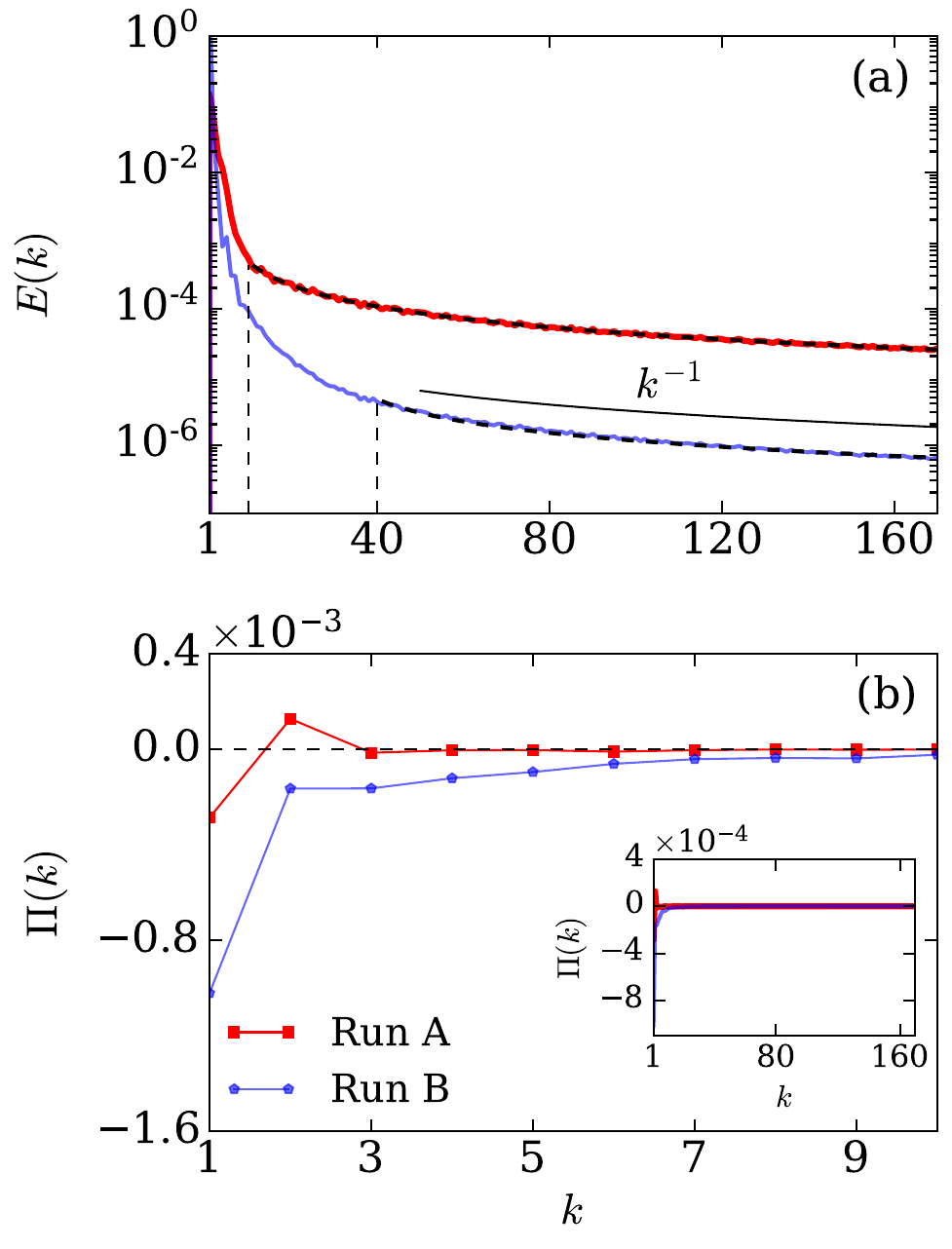} 
	\caption{(Color online) For the  Runs A, B of 2D Euler turbulence: plots of the averaged energy spectra, $ E(k)  $, and fluxes, $ \Pi(k) $, of the asymptotic states. Figure (b) exhibits $ \Pi(k) $ for small $ k $'s, while inset shows $ \Pi(k) $ for the whole range.
	}\label{fig:2D_Ek}
\end{figure}

In Fig.~\ref{fig:2D_Ek}(a,b), we plot the averaged energy spectra and fluxes of the steady states of  the two runs. The two runs have the following energy spectra:
\bea
E(k) & = & \frac{k}{-2518+237 k^2}~	\mathrm{Run A: for}~k > 10, \label{eq:RunA} \\
E(k) & = & \frac{k}{-6357840+9361 k^2}~\mathrm{Run B: for}~k > 40, \label{eq:RunB} \nonumber \\
\eea
For Runs A and B, $ \sqrt{\Omega/E} = 15.8, 1.2$. Since $\sqrt{\Omega/E} \le k_\Omega$ (centroid of enstrophy),  the  wavenumbers far beyond $  k_\Omega $ are dominated by enstrophy leading to $ E(k) \propto k^{-1} $, which corresponds to an equipartition of enstrophy~\cite{Nazarenko:book:WT,Verma:arxiv2022}.

The average energy flux $ \la \Pi(k)  \ra \approx 0$ for intermediate and large $ k $'s. But, $ \la \Pi(k)  \ra < 0$ for small $ k $'s. In particular,   $ \min [\la \Pi(k)  \ra] \approx -3 \times10^{-4}, -10^{-3}$ for Runs A and B respectively. Thus, the nonzero $ \Pi(k) $ breaks the \textit{detailed balance of energy transfers}. Hence, 2D Euler turbulence is out of equilibrium. See \citet{Verma:arxiv2022} for more details.

\begin{figure}
	\includegraphics[width=8.5cm]{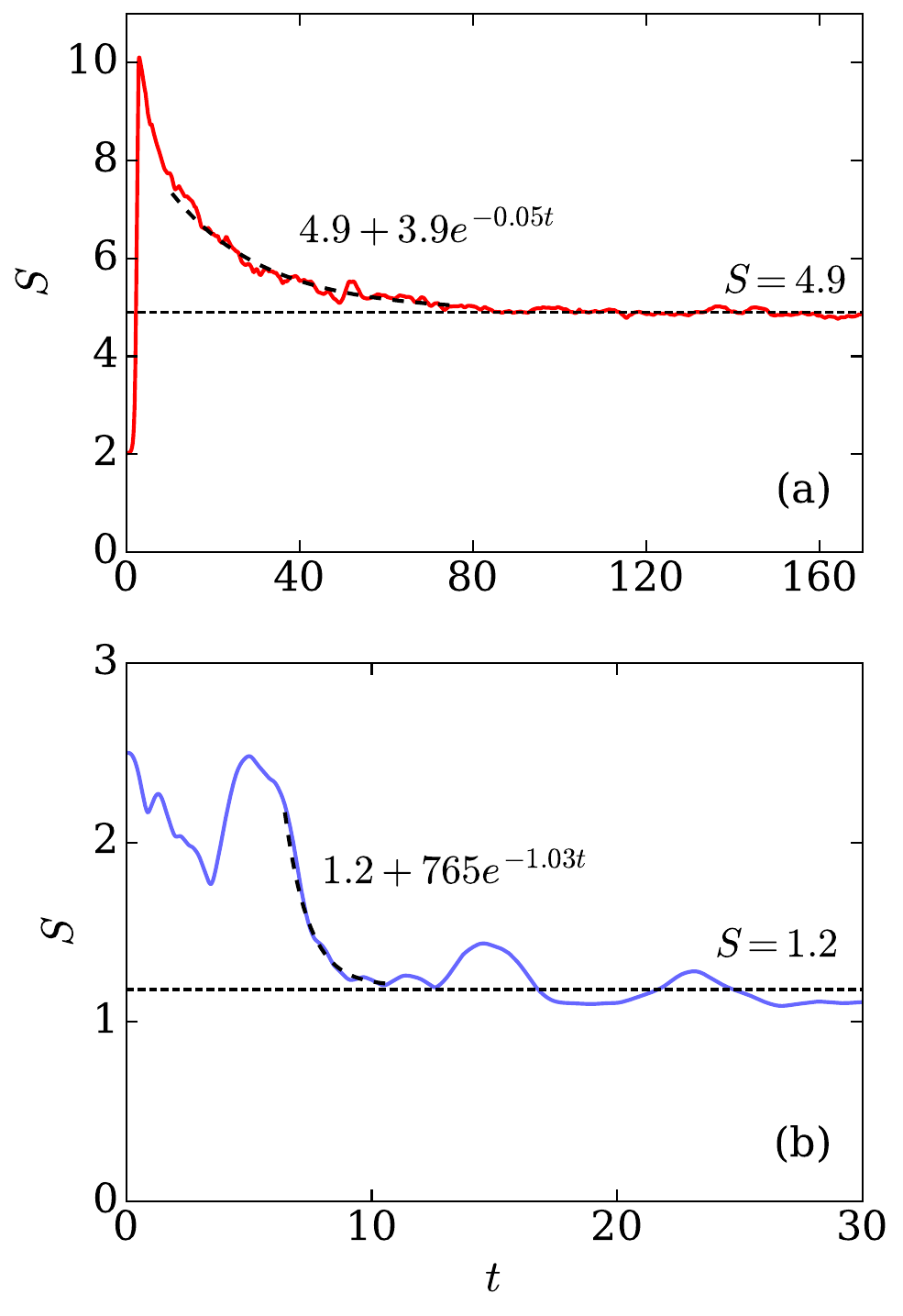} 
	\caption{(Color online) Plots of the temporal evolution of hydrodynamic entropies  for Runs A, B (a, b) of 2D Euler turbulence respectively.
	In each case, after initial transients, the entropy decreases with time and asymptotes to an approximate constant value.
	}\label{fig:2D_entropy}
\end{figure}

\citet{Verma:arxiv2022} computed the hydrodynamic  entropies of the  2D Euler flows of Runs A and B. The plots of   the entropy time series are shown in Fig.~\ref{fig:2D_entropy}.  For Runs A and B, after initial fluctuations, the entropies decrease exponentially to asymptotic values of 4.9 and 1.2 respectively.  These values are smaller than the maximum possible value, which is $ \log_2(M) \approx 16.5$, where $ M \approx \pi (512/3)^2 $.  Thus, \citet{Verma:arxiv2022} showed that the hydrodynamic entropy of 2D Euler turbulence, an isolated system, decreases  with time for a significant duration.  Hence, 2D Euler turbulence is a rare isolated  system that exhibits evolution from \textit{disorder to order}.

\section{DISCUSSIONS and CONCLUSIONS}
 \label{sec:conclusions}
In this paper, we review the equilibrium and nonequilibrium properties of 2D and 3D Euler turbulence. For $\delta$-correlated velocity field as an initial condition, both 2D and 3D Euler turbulence exhibit equilibrium behaviour, predicted  by \citet{Lee:QAM1952} and \citet{Kraichnan:JFM1973}.  However, for ordered initial condition, 3D Euler turbulence evolves from order to disorder. The above thermalization of 3D Euler turbulence follows a generic path for energy conserving system. Here, the nonlinear energy transfer $T(k,t) \rightarrow 0$ as $t\rightarrow \infty$. The energy flux too vanishes asymptotically. 

The thermodynamic entropy of Euler turbulence remains constant. Hence,  \citet{Verma:arxiv2022} constructed hydrodynamic entropy to quantify the variations in order of Euler turbulence. They showed   that the hydrodynamic entropy of 3D Euler turbulence increases monotonically with time.

\begin{figure}
	\includegraphics[scale=0.9]{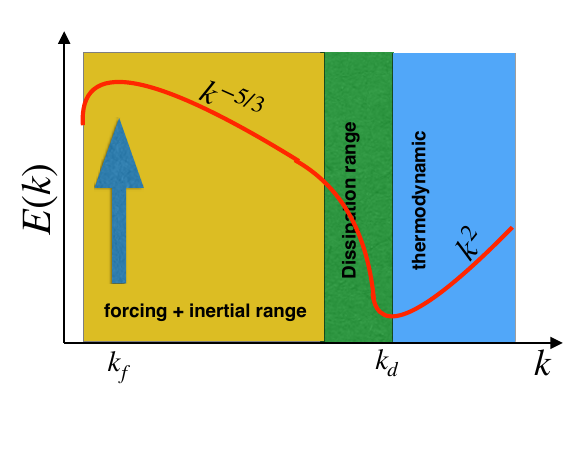}
	\caption{A schematic diagram exhibiting the  energy spectrum $E(k)$ of 3D forced hydrodynamic turbulence (includes viscous effects).  The yellow and green regions  represent the forcing, inertial, and dissipation ranges of hydrodynamic description, while the blue region represents the thermal motion of molecules.  The former regions are out of equilibrium, while the latter one is in quasi-equilibrium.    }
	\label{fig:NS}
\end{figure}

Interestingly, we can extrapolate the above thermalization process to Navier-Stokes equation that includes viscous dissipation, as well as external forcing at large scales.    A fluid is composed of molecules whose total energy is conserved.  However, we can separate the system into two parts: (a) coherent flow, which is described by the flow equation under continuum approximation, and (b) random or thermal motion of the molecules in the co-moving frame of the flow.  In Kolmogorov's picture of turbulence, hydrodynamic range of scales includes forcing, inertial, and dissipation range, while the random motion of molecules is described by thermodynamics.  These scales are exhibited by different colors in Fig.~\ref{fig:NS}. The transition wavenumber between the hydrodynamic and thermodynamic ranges may be approximated by the Kolmogorov wavenumber, $k_d = (\epsilon/\nu^3)^{1/4}$, where $\nu$ is the kinematic viscosity.   The above picture is similar to findings in  recent works by Shukla et al.~\cite{Shukla:PRE2019} and \citet{Eyink:PRE2022}.

The above arguments can be extended to quantum systems, at least to superfluids and Bose-Einstein gas.   Many experiments and  numerical simulations of such systems yield Kolmogorov-like $k^{-5/3}$ spectrum~(\citet{Krstulovic:PRL2011}, \citet{Madeira:ARCMP2020}, \citet{Fonda:PNAS2019}, \citet{Shukla:PRA2019},  \citet{Skrbek:PF2012},
and references therein) that requires dissipation at small scales.  Small-scale dissipation in such systems are attributed to interactions of condensate with thermal clouds, or to  decay of vortical motion into phonon excitations~(\citet{Barenghi:PNAS2014} and references therein).    This feature may appear odd because quantum systems are energy conserving.  But, the  multiscale energy transfer in Euler turbulence provides an interesting framework to introduce quantum dissipation and thermalization~\cite{Caldera:AP1983,Mohsen:book,Weiss:book:QuantumDiss}.    This framework could be an alternative to  other approaches that are typically based on modeling the   interactions between the system and the heat bath~\cite{Mohsen:book,Weiss:book:QuantumDiss,DAlessio:AdvP2016}.     

In contrast, 2D Euler turbulence remains out of equilibrium, at least for several ordered initial conditions. This is contrary to what we expect for a conservative system with a  large degree of freedom. Here, the nonlinear energy transfer and energy flux do not vanish asymptotically. Thus, for such a scenario, the final states of 2D Euler turbulence are not stationary. Interestingly, the hydrodynamic entropy of 2D Euler turbulence decreases in the asymptotic regime.  Thus, 2D Euler turbulence is a unique \textit{isolated} system that exhibits evolution form disorder to order.   

\begin{acknowledgments}
The {authors thank}  Arul Lakshminarayan, Stephan Fauve, Marc Brachet, Alex Alexakis, Hal Takasi, Anurag Gupta, Saikat Ghosh, Franck Plunian,  Rodion Stepanov, and Giorgio Krstulovic  for useful  discussions.   We also thank the organizers of APPC15 for hosting the meeting. This work is supported by the project 6104-1  from the Indo-French Centre for the Promotion of Advanced Research (IFCPAR/CEFIPRA).  Soumyadeep Chatterjee is supported by INSPIRE fellowship (IF180094) from Department of Science \& Technology, India.

\end{acknowledgments}

\appendix
\section{Modified PEFRL algorithm} 
\label{sec:appendix}

\begin{figure}
	\includegraphics[width=8.5cm]{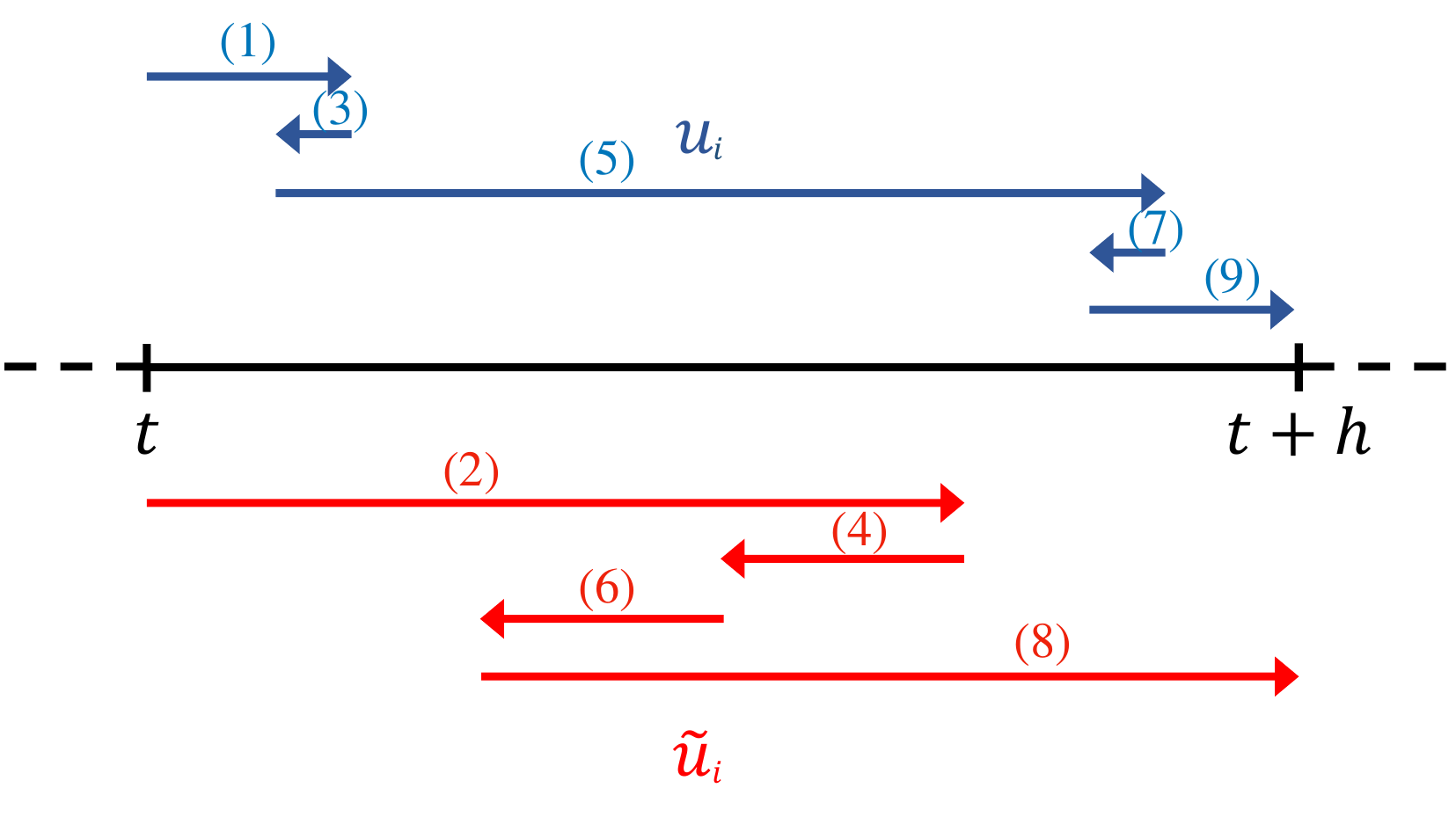} 
	\caption{(Color online) Schematic diagram of modified PEFRL algorithm.
	}\label{fig:PEFRL}
\end{figure}

\citet{Omelyan:CPC2002} extended Forest-Ruth (FR) algorithm~\cite{Forest:PD1990} to solve differential equations associated with conservative systems.  Their scheme is called \textit{position-extended Forest-Ruth-like} (PEFRL). The PEFRL algorithm was devised for molecular dynamics simulations. We modify it to  simulate Euler flow. The evolution of velocity of an inviscid fluid is an explicit function of $ {\bf u} $. In the following, we write equations for $u_i$,   the $i$th component of velocity:
\be
\frac{du_i}{dt} = R(u_i). \label{eq:system}
\ee
Various steps involved in  time advance from $t$ to $t+h$ are as follows (see Fig.~\ref{fig:PEFRL}):
\bea
&(1)& u_i  = u_i(t) + \xi hR(u_i(t)), \nonumber \\
&(2)& \widetilde{u_i} = u_i(t) + (1-2\lambda) \frac{h}{2} R(u_i), \nonumber \\
&(3)& u_i  = u_i + \chi hR(\widetilde{u_i}), \nonumber \\
&(4)& \widetilde{u_i} = \widetilde{u_i} + \lambda hR(u_i), \nonumber \\
&(5)& u_i  = u_i + (1-2(\chi+\xi))hR(\widetilde{u_i}), \nonumber \\
&(6)& \widetilde{u_i} = \widetilde{u_i} + \lambda hR(u_i), \nonumber \\
&(7)& u_i  = u_i + \chi hR(\widetilde{u_i}), \nonumber\\
&(8)& \widetilde{u_i} = \widetilde{u_i} + (1-2\lambda) \frac{h}{2} R(u_i), \nonumber\\
&(9)& u_i(t+h) = u_i + \xi hR(\widetilde{u_i}). 
\label{eq:PEFRL}
\eea
Here, $\widetilde{u}$ is an intermediate velocity placeholder, whereas $\xi$, $\lambda$, and $\chi$ are constants. \citet{Omelyan:CPC2002} computed the optimized values of these constants as
\begin{eqnarray}
\xi &=& 0.1786178958448091 \nonumber\\
\lambda &=& -0.2123418310626054 \nonumber \\
\chi &=& -0.06626458266981849. \label{eq:PEFRL_constants}
\end{eqnarray} 

%



\end{document}